\documentclass[%
 reprint,
superscriptaddress,
preprintnumbers,
 amsmath,amssymb,
 aps,
]{revtex4-1}

\usepackage{graphicx}
\usepackage{hyperref}
\usepackage{acro}

\usepackage{multirow}
\usepackage[version=4]{mhchem}
\usepackage{dsfont}
\usepackage{float}
\usepackage{caption}
\usepackage{subcaption}
\usepackage{braket}
\usepackage{color}

\hypersetup{
	colorlinks   = true,
	citecolor    = blue,
	linkcolor = blue,
    urlcolor = blue
}

\begin{document}

\title{Many-Body Destabilization of Intermediate Oxygen-Hole States}

\author{Anirudh Adavi}
\thanks{These authors contributed equally.}
\affiliation{Department of Materials Science and Engineering,
Massachusetts Institute of Technology, Cambridge, MA 02139, USA}

\author{Kayahan Saritas}
\thanks{These authors contributed equally.}
\affiliation{Materials Sciences and Technology Division,
Oak Ridge National Laboratory, Oak Ridge, TN 37831, USA}
\thanks{This manuscript has been authored by UT-Battelle, LLC under Contract No.\ DE-AC05-00OR22725 with the U.S.\ Department of Energy. The United States Government retains and the publisher, by accepting the article for publication, acknowledges that the United States Government retains a non-exclusive, paid-up, irrevocable, worldwide license to publish or reproduce the published form of this manuscript, or allow others to do so, for United States Government purposes. The Department of Energy will provide public access to these results of federally sponsored research in accordance with the DOE Public Access Plan (\url{https://www.energy.gov/doe-public-access-plan}).}

\author{Ming Lei}
\affiliation{Department of Materials Science and Engineering,
Massachusetts Institute of Technology, Cambridge, MA 02139, USA}

\author{Das Pemmaraju}
\affiliation{IBM Research, 555 Bailey Ave, San Jose, CA 95141-1003, USA}

\author{Paul R. C. Kent}
\affiliation{Computational Sciences and Engineering Division,
Oak Ridge National Laboratory, Oak Ridge, TN 37831, USA}

\author{Iwnetim I. Abate}
\email[Corresponding author: ]{iabate@mit.edu}
\affiliation{Department of Materials Science and Engineering,
Massachusetts Institute of Technology, Cambridge, MA 02139, USA}

\begin{abstract}

Oxygen holes in transition-metal oxides can appear as localized polarons, symmetry-delocalized ligand holes, or intermediate states whose stability is controlled by subtle electron-correlation effects. In layered Na$_{2-x}$Mn$_3$O$_7$, hybrid density functional theory (DFT) predicts an unusual bond-centered split oxygen-hole polaron stabilized near ordered Mn vacancies. Here we resolve the nature of this state using diffusion Quantum Monte Carlo (QMC). Although hybrid DFT favors the split configuration, QMC reverses the energetic ordering and identifies the localized oxygen polaron as the lower-energy state. The result is robust to the class of trial wavefunctions used, including hybrid and generalized-gradient DFT wavefunctions. Many-body spin densities further show that the nominal split state partially collapses toward a localized polaron. Because localized and split configurations produce similar O K-edge spectral features, this qualitative failure is not resolved by conventional X-ray absorption signatures alone. These findings identify Na$_{2-x}$Mn$_3$O$_7$ as a stringent benchmark for oxygen-hole polarons and reveal a failure mode of hybrid functionals in correlated oxides.

\end{abstract}

\maketitle

Small polarons are central to the electronic structure of transition-metal oxides, where strong coupling among lattice, charge, and spin degrees of freedom can localize carriers and govern transport, magnetism, optical response, and redox activity\cite{Coropceanu, huang2017jahn, nagaev_colossal, franchini_polarons_2021, PhysRevB.54.R9592, 2D_perovskites, LARSSON200335, Luong_pccp, reticcioli2020small}. In many correlated oxides, the competition between localized and itinerant electronic behavior determines emergent phenomena ranging from superconductivity\cite{KELLER200838, trugman2004jahn, condmat7010010, bednorz1986possible, m1999large, N_F_Mott_1993} and charge ordering\cite{savitzky2017bending, li2014direct} to colossal magnetoresistance\cite{nagaev_colossal, ramirez_colossal_1997} and multiferroicity\cite{PhysRevB.82.140101,xu2019electron, miyata_large_2017, miyata_ferroelectric_2018}. Oxygen-hole polarons are particularly subtle because holes introduced on oxygen can compete among several limiting descriptions: localized O$^-$ states, symmetry-delocalized ligand-hole states, and intermediate configurations stabilized by local bonding environments and defect chemistry\cite{schirmer_o_2006}. The energetic separation between these competing electronic states is often small, making their description highly sensitive to electron correlation and lattice relaxation.

This problem is especially relevant in transition-metal oxides exhibiting oxygen-hole formation and correlated ligand states. In such systems, hole doping can drive the emergence of localized oxygen polarons, symmetry-delocalized ligand holes, or intermediate bonding configurations depending on the competition among covalency, lattice relaxation, and Coulomb interactions\cite{schirmer_o_2006, abate_coulombically-stabilized_2021, YU1995233, E_Possenriede_1992, house2023delocalized, wang2023stabilizing}. Distinguishing among these competing electronic states is central to understanding charge localization, defect-driven symmetry breaking, and correlated electronic behavior in complex oxides. At the same time, oxygen-hole systems provide a stringent test for electronic-structure methods because localized and partially delocalized states can differ by only small energy scales while exhibiting qualitatively different real-space electronic structures\cite{schirmer_o_2006}.

Density functional theory (DFT)\cite{PhysRev.136.B864, PhysRev.140.A1133} has become the dominant framework for investigating these problems because it provides simultaneous access to electronic structure and lattice relaxation in complex solids. However, approximate exchange-correlation functionals often struggle to describe localized electronic states. Semilocal functionals\cite{PhysRevLett.77.3865} tend to over-delocalize holes because of self-interaction error, whereas methods such as DFT+$U$\cite{PhysRevB.44.943} or hybrid functionals\cite{10.1063/1.1564060} can over-stabilize symmetry-broken localized solutions depending on the treatment of exchange and screening. Systems containing competing localization motifs therefore provide particularly stringent benchmarks for approximate electronic-structure methods. Explicit many-body approaches are especially valuable in such cases because they provide a route to determine whether predicted intermediate oxygen-hole states are genuine correlated ground states or artifacts of the approximate treatment of exchange and correlation.

\begin{figure*}[t]
\centerline{
\includegraphics[width=0.6\textwidth]{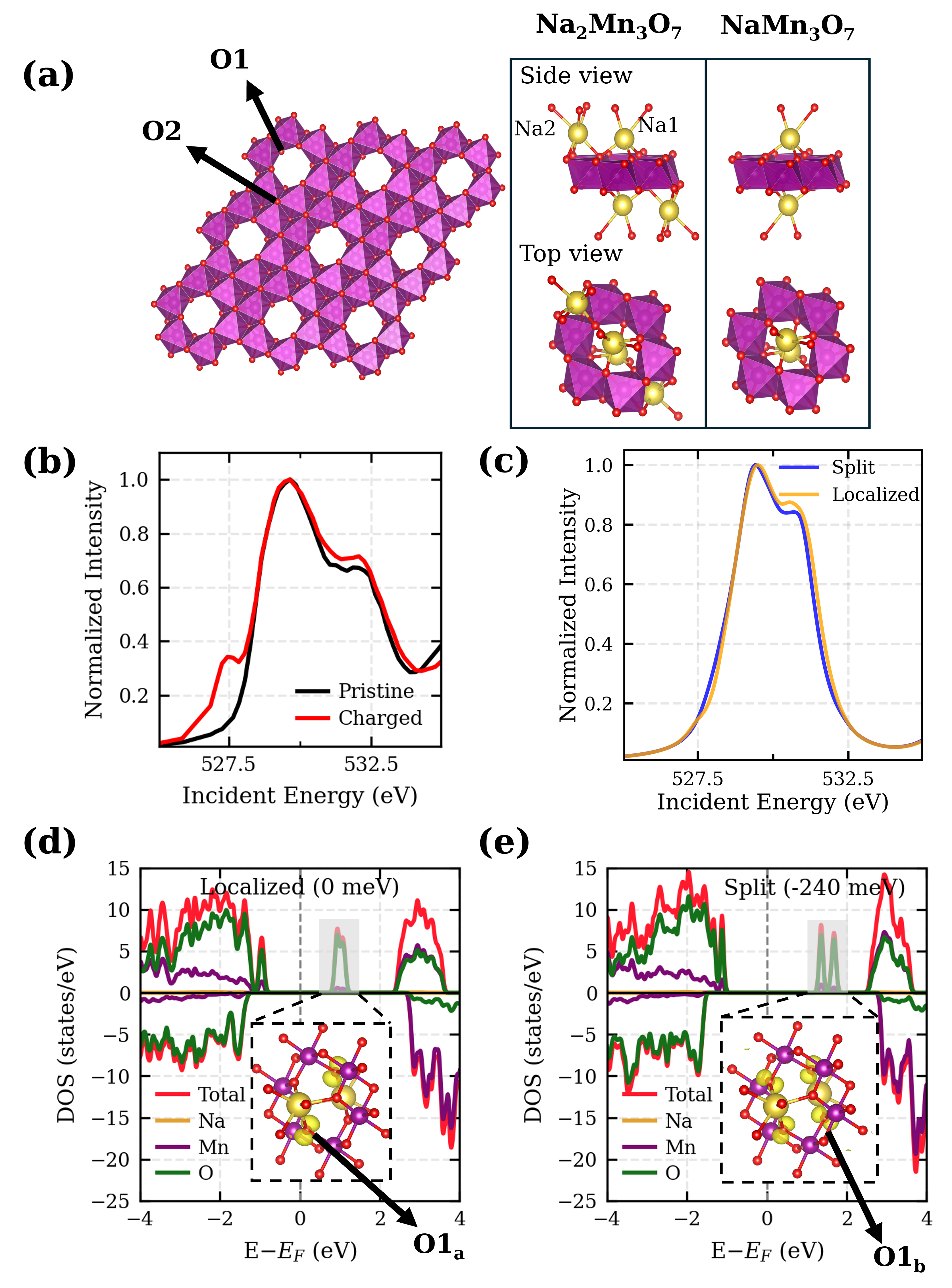}
}
\caption{Physical origin and spectroscopic ambiguity of oxygen-hole states in Na$_{2-x}$Mn$_3$O$_7$.
(a) Crystal structure of Na$_{2-x}$Mn$_3$O$_7$ showing the ordered Mn-vacancy environment and oxygen sublattices relevant to oxygen-hole formation. Oxygen atoms neighboring the Mn vacancy provide the preferred sites for oxygen-hole localization after Na removal. Na atoms located beneath the Mn vacancy layer are labeled Na1, and they occupy trigonal prismatic sites in the layers above and below the Mn vacancies. Na atoms situated farther from the vacancy layer are labeled Na2 and are octahedrally coordinated. Oxygen anions bonded to two Mn atoms are labeled as O1, and those bonded to three Mn atoms are labeled as O2.
(b) Experimental O K-edge fluorescence-yield spectra showing emergence of the oxygen-hole pre-edge feature near 527.5~eV upon charging, consistent with ligand-hole formation.
(c) Calculated O K-edge spectra for localized and split-polaron configurations. Both configurations produce qualitatively similar spectra despite their substantially different real-space electronic structures, and neither reproduces the experimentally observed pre-edge feature well.
(d) (e) DOS/LDOS distributions for localized and split-polaron states illustrating the contrasting oxygen-hole localization motifs and oxygen-derived mid-gap states associated with hole formation.
}
\label{fig:XAS_DFT}
\end{figure*}

Layered Na$_{2-x}$Mn$_3$O$_7$ provides an ideal model system in which to examine this competition between localization, hybridization, and many-body correlation effects\cite{abate_coulombically-stabilized_2021, acs.chemmater.7b01390, acs.chemmater.9b00772, aenm.201800409}. The material consists of edge-sharing MnO$_6$ octahedral layers separated by Na layers, with ordered Mn vacancies forming a maple-leaf lattice in the transition-metal plane [Fig.~\ref{fig:XAS_DFT}(a)]. Previous experimental studies identified Na$_{2-x}$Mn$_3$O$_7$ as an unusual oxygen-hole in which Na removal produces stable, reversible ligand-hole states\cite{abate_coulombically-stabilized_2021}. Resonant inelastic X-ray scattering (RIXS) and O K-edge fluorescence-yield spectra showed the emergence of oxygen-derived pre-edge intensity near 527.5 eV upon deintercalation, consistent with the formation of oxygen-hole states [Fig.~\ref{fig:XAS_DFT}(b)]\cite{abate_coulombically-stabilized_2021}. Earlier electronic-structure calculations further suggested that these holes form oxygen-centered polarons stabilized by the local defect environment surrounding the ordered Mn vacancies\cite{abate_coulombically-stabilized_2021}.

In NaMn$_3$O$_7$, removal of one Na atom per formula unit creates two oxygen holes per triclinic unit cell. The ordered Mn vacancies act as negatively charged acceptor centers that favor hole localization on neighboring oxygen atoms\cite{schirmer_o_2006}. Oxygen atoms bordering the Mn vacancy (O1 in Fig. \ref{fig:XAS_DFT}(a)) are therefore electronically distinct from the remaining oxygen sublattice and provide the preferred sites for oxygen-hole formation\cite{abate_coulombically-stabilized_2021}. Conventional DFT+$U$ calculations using the generalized-gradient approximate (GGA) functional PBE\cite{PhysRevLett.77.3865} predict localized oxygen polarons confined to two individual oxygen atoms neighboring the Mn vacancy (O1\textsubscript{a} in Fig. \ref{fig:XAS_DFT}(d)). In contrast, hybrid DFT calculations using the HSE functional\cite{10.1063/1.1564060} predict a qualitatively different split-hole polaron in which the hole density is shared across neighboring oxygen atoms\cite{abate_coulombically-stabilized_2021} (O1\textsubscript{a} and O1\textsubscript{b} sites). Importantly, this HSE stabilization persists even when symmetry-broken localized oxygen-hole configurations are initialized, indicating that the split state is an energetically preferred HSE solution rather than a consequence of constrained symmetry. Such partial delocalization is highly unusual in correlated oxides, where oxygen-hole states are typically expected to undergo either strong local symmetry breaking or delocalization into extended ligand-hole bands\cite{house2023delocalized, radin_manganese_2019}.

The stabilization of such a split-polaron state raises a central question: does Na$_{2-x}$Mn$_3$O$_7$ host a genuine bond-centered oxygen-hole polaron, or does hybrid DFT qualitatively mis-rank competing localization motifs? The answer is nontrivial because the system contains several competing physical effects. Charge localization favors local symmetry breaking and lattice distortion to form oxygen-centered polarons. Coulomb interactions near the Na-vacancy environment can instead stabilize more distributed charge configurations. Hybridization between neighboring oxygen orbitals further modifies the balance between localization and delocalization. The resulting energetic competition occurs on relatively small energy scales and is therefore highly sensitive to the treatment of electron correlation.

Spectroscopy alone does not resolve this ambiguity. Calculated O K-edge spectra for localized and split-polaron configurations are not qualitatively distinguishable despite their substantially different real-space charge distributions [Fig.~\ref{fig:XAS_DFT}(c)]\cite{abate_coulombically-stabilized_2021}. Thus, conventional X-ray absorption signatures cannot uniquely identify the microscopic nature of the oxygen-hole state. The contrasting localized and split oxygen-hole distributions obtained from the local density of states (LDOS) analysis are illustrated in Fig.~\ref{fig:XAS_DFT}(d,e). Determining the true ground-state configuration therefore requires an explicit comparison of the energetics and many-body electronic structure of the competing localized and split-polaron states.

To resolve this problem, we performed fixed-node diffusion Quantum Monte Carlo (DMC)\cite{10.1063/1.431514, PhysRevLett.45.566, Kim_2018, kent_qmcpack_2020} calculations using trial wavefunctions generated from DFT calculations using both the hybrid functional HSE and the GGA functional PBE with $U$+$V$\cite{CAM2010} corrections. DMC provides an explicitly many-body treatment of electron correlation and is particularly valuable in systems where competing localization motifs differ by small energy scales\cite{FOU2001}. In the present work, we combine hybrid DFT, PBE+$U$+$V$ DFT, variational Monte Carlo (VMC), and DMC calculations to determine both the energetic ordering and physical character of the competing oxygen-hole states in NaMn$_3$O$_7$.

\begin{figure*}[ht]
\centerline{
\includegraphics[width=0.9\textwidth]{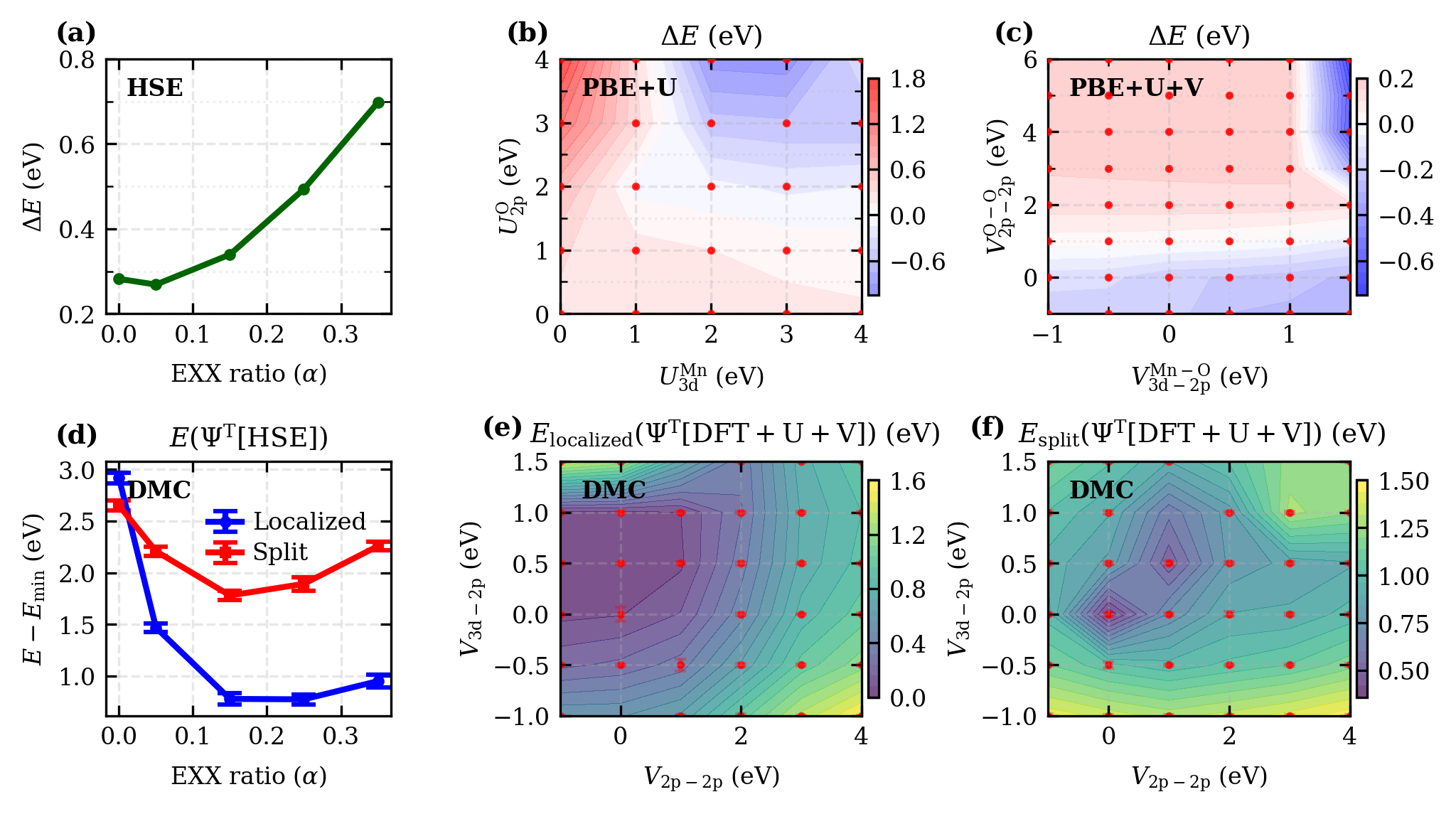}
}
\caption{Many-body reversal of the split-polaron stabilization predicted by hybrid DFT.
(a) Relative energy difference between localized and split-polaron configurations ($\Delta E = E_{\mathrm{localized}} - E_{\mathrm{split}}$) within DFT [HSE] as a function of exact-exchange fraction~$\alpha$. Hybrid DFT systematically stabilizes the split-polaron state across the explored parameter range.
(b) (c) DFT [PBE] energy differences across grids of tested (b) $U$ values and (c) $V$ values. In both cases, a crossover in the energy difference is observed.
(d) DMC energies obtained using HSE-generated trial wavefunctions. $E_{\min}$ is the minimum total energy obtained across the DMC calculations in (d) through (f). Although HSE favors the split-polaron state near the optimal exact-exchange fraction ($\Delta E > 0$), DMC reverses the energetic ordering and stabilizes the localized oxygen hole polaron ($\Delta E < 0$).
(e) (f) DMC total energies of localized and split-polaron configurations using PBE+$U$+$V$-generated trial wavefunctions. Explicit many-body treatment stabilizes the localized oxygen polaron despite substantial differences in the underlying approximate DFT descriptions.
}
\label{fig:reversal}
\end{figure*}

Finite-size\cite{PhysRevB.78.125106} errors were minimized using twist averaging\cite{Lin2001,Annaberdiyev2024} and supercell extrapolation procedures. Nodal surface errors, the main limitation of accuracy in DMC calculations\cite{Kim_2018, FOU2001}, were controlled by parameter optimization in DFT-based trial wavefunctions (more details can be found in Supplemental Material Section VI). Trial wavefunctions were generated using both HSE and PBE+$U$+$V$ calculations in order to assess whether the many-body energetic ordering depends strongly on the underlying approximate single-particle description.

We first examine the behavior of the approximate DFT descriptions. In hybrid DFT, the split-polaron configuration is lower in energy than the localized-polaron configuration across the range of exact-exchange fractions investigated [Fig.~\ref{fig:reversal}(a)]. Increasing the exact-exchange fraction further stabilizes the split-polaron state relative to the localized state. The corresponding electronic structure contains an oxygen-derived unoccupied mid-gap state associated with hole formation. Integration of the local density of states around the mid-gap feature reveals a hole-state density distributed across four oxygen atoms neighboring the Mn vacancy, corresponding to two split-hole polarons, as in Fig.~\ref{fig:XAS_DFT}(e). Additional projected and local densities of states are provided in Supplemental Material Fig. S1.

The behavior within PBE+$U$ differs substantially from HSE. Although PBE+$U$ can produce oxygen-centered hole states and mid-gap features similar to those obtained from hybrid DFT, the resulting local densities remain considerably more localized. Even nominal split-polaron geometries tend to collapse toward partially localized oxygen-hole states. This tendency indicates that onsite Hubbard interactions alone are insufficient to stabilize the fully distributed split-polaron density obtained within HSE. Including inter-site interactions within DFT+$U$+$V$, particularly the O-2$p$/O-2$p$ interaction near the Mn-vacancy environment, partially restores the split-polaron character predicted by HSE, demonstrating that the stability of the state depends sensitively on inter-site hybridization and Coulomb interactions. Representative PBE+$U$ and PBE+$U$+$V$ parameter scans are shown in Supplemental Material Figs. S2–S3.

Despite the qualitative differences among the DFT approximations, the DMC calculations consistently reverse the energetic ordering predicted by HSE, as summarized in Fig.~\ref{fig:reversal}. Using HSE-generated trial wavefunctions, the DMC total energies for both localized and split configurations are minimized near an exact-exchange fraction of approximately $\alpha$ $\approx$ ~0.15. At this value, hybrid DFT favors the split-polaron state by approximately 0.28~eV [Fig.~\ref{fig:reversal}(a)]. DMC instead predicts the localized-polaron configuration to be lower in energy by approximately 1~eV [Fig.~\ref{fig:reversal}(d)]. Thus, explicit many-body treatment destabilizes the split-polaron state and reverses the energetic ordering by more than 1~eV relative to HSE.

Additional calculations confirm that this reversal is not an artifact of finite-size effects. Finite-size extrapolation is summarized in Supplemental Material Fig. S5. Although the absolute DMC energies vary slightly with supercell size and twist averaging, the energetic preference for localization remains robust. The disagreement between hybrid DFT and DMC therefore reflects a genuine difference in the predicted correlated ground state.

We next examined whether this conclusion depends strongly on the choice of trial wavefunction. The comparison between HSE- and PBE+$U$+$V$-generated trial wavefunctions is summarized in Fig.~\ref{fig:reversal}. DFT calculations show a crossover region in the energy difference between the localized and split configurations with a GGA description, as illustrated in Fig.~\ref{fig:reversal}(b,c). The potential energy wells for DMC wavefunctions with PBE+$U$+$V$ trial wavefunctions are shown in Fig.~\ref{fig:reversal}(e,f), showing a lower minimum energy across the V parameter scan for the localized state. DMC calculations based on both classes of trial wavefunctions consistently favor the localized oxygen polaron, even though the underlying approximate DFT descriptions differ substantially. The magnitude of the energy difference between localized and split configurations decreases somewhat for DFT+$U$+$V$-generated trial states, but the energetic ordering remains unchanged. The many-body preference for localization is therefore robust across the tested HSE- and DFT+$U$+$V$-generated trial wavefunctions.

\begin{figure*}[ht]
\centerline{
\includegraphics[width=0.8\textwidth]{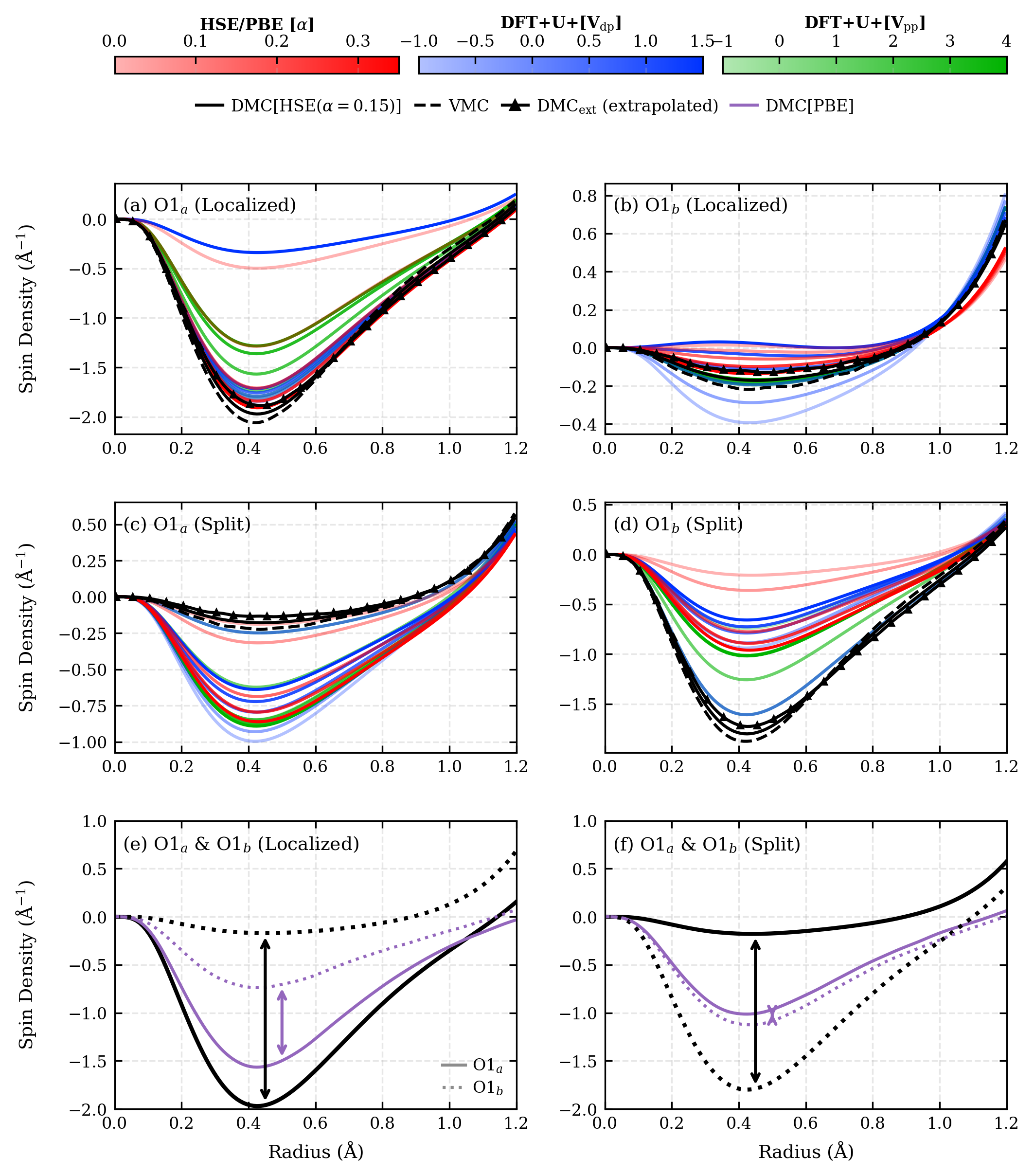}
}
\caption{Many-body destabilization of the split-polaron state.
(a,b) Radial spin-density distributions for localized oxygen polarons showing robust localization across DFT, VMC, and DMC descriptions. The spin polarization remains concentrated predominantly on the O1\textsubscript{a} sites neighboring the Mn vacancy, except when approaching the GGA limit.
(c,d) Radial spin-density distributions for split oxygen polarons showing qualitative disagreement between DFT and DMC descriptions. Most DFT spin densities for the split-polaron configuration show nearly symmetric spin polarization distributed across neighboring oxygen atoms surrounding the Mn vacancy, while DMC[HSE] and some DFT+$U$+$V$ predicts strongly enhanced asymmetry, with spin density localizing strongly on a different O1 site neighboring the Mn vacancy.
(e, f) Trial wavefunction dependence of DMC oxygen spin density at O1\textsubscript{a}/O1\textsubscript{b} sites.  Spin-density asymmetry between DMC[HSE] and DMC[PBE] showing weaker asymmetry of DMC[PBE] wavefunctions but demonstrating a qualitatively similar result. Explicit many-body treatment enhances the localization tendency, although the degree of split-polaron collapse is sensitive to the underlying trial wavefunction.
}
\label{fig:destabilization}
\end{figure*}

The many-body spin densities provide further insight into the physical origin of the instability of the split-polaron state, as shown in Fig.~\ref{fig:destabilization}. Consistent with the earlier discussion of the XAS results, total charge densities show only modest differences among the various DFT approximations and the DMC calculations, despite the distinct localization motifs of the competing oxygen-hole states [Fig. S8]. This helps explain why conventional X-ray absorption spectroscopy cannot uniquely resolve the precise localization character of the oxygen-hole states shown in Fig.~\ref{fig:XAS_DFT}. In contrast, spin-density distributions are considerably more sensitive to the underlying localization physics because the oxygen hole corresponds to an unpaired spin\cite{acs.chemmater.9b00772}. In the localized-polaron state, the spin density is concentrated predominantly on a single oxygen atom, whereas in the split-polaron configuration the spin polarization is distributed more symmetrically across neighboring oxygen sites. Spin polarization densities therefore provide a clearer fingerprint of the underlying polaron character. Additional radial spin and charge-density comparisons are provided in Supplemental Material Fig. S6-8.

For the localized-polaron configuration, the DMC spin density remains concentrated on the O1\textsubscript{a} oxygen sites neighboring the Mn vacancy that host the oxygen holes within DFT [Fig.~\ref{fig:destabilization}(a,b)]. Relative to the DFT descriptions, the DMC calculations generally enhance the degree of spin polarization on these oxygen sites, indicating a stronger preference for symmetry breaking under explicit many-body treatment. The qualitative character of the localized-polaron state nevertheless remains consistent across DFT and DMC.

The split-polaron configuration behaves very differently. Within DFT[HSE], the spin polarization is nearly symmetrically distributed across neighboring oxygen atoms surrounding the Mn vacancy [Fig.~\ref{fig:destabilization}(c,d)]. For the HSE-generated split-polaron trial state, DMC strongly breaks this balance. One oxygen site develops substantially larger spin polarization, while the neighboring oxygen site becomes comparatively weakly polarized. The resulting many-body spin-density profile therefore more closely resembles a localized oxygen polaron rather than a genuinely stable split-polaron state. This behavior is most pronounced for the DMC with HSE trial wavefunctions (DMC[HSE]) description; DMC calculations based on PBE trial wavefunctions exhibit a much weaker asymmetry [Fig.~\ref{fig:destabilization}(e,f)], although the overall energetic preference for localization remains unchanged.

This behavior demonstrates that the many-body wavefunction does more than alter the energetic ordering between localized and split configurations; it can qualitatively change the physical character of the nominal split-polaron solution itself. Within the single-determinant DMC[HSE] framework used here, the many-body projection partially collapses the HSE split-polaron density toward localized oxygen-hole configurations. In practice, the system at the DMC-optimized $\alpha$ does not realize distinct localized and split states but instead yields two localized variants in which the hole density resides predominantly on one oxygen site or the other.

The observed collapse likely reflects enhanced static correlation and near-degeneracy among competing oxygen-hole symmetry-broken states. Hybrid DFT predicts nearly degenerate oxygen-hole distributions across neighboring oxygen sites, suggesting that multiple symmetry-related localization motifs lie close in energy. In such situations, single-determinant trial wavefunctions may not provide a sufficiently flexible description of the correlated many-body state. More sophisticated multideterminant or orbital-optimization approaches could potentially stabilize a more balanced distribution of the oxygen-hole density. More complete radial spin-density comparisons between DFT, VMC, and DMC are provided in Supplemental Material Figs. S6-S8. Nevertheless, within the present many-body framework, the split-polaron state is not robust relative to localization.

These results have broader implications for correlated oxygen-hole states in transition-metal oxides and other systems exhibiting competing localization and hybridization phenomena. Hybrid functionals are widely used to study correlated oxygen-hole states because they partially correct the self-interaction error present in semilocal DFT. The present results demonstrate, however, that hybrid functionals can qualitatively misrepresent competing oxygen-hole localization motifs when their stability depends on subtle balances among localization, hybridization, and Coulomb interactions. In Na$_{2-x}$Mn$_3$O$_7$, HSE stabilizes a spectroscopically plausible intermediate oxygen-hole state that becomes unstable under explicit many-body treatment.

More broadly, the work highlights the limitations of relying exclusively on agreement with a limited set of spectroscopic features to validate correlated electronic states predicted by approximate electronic-structure methods. Both localized and split-polaron configurations produce similar calculated spectra to experimentally observed O K-edge features despite their substantially different real-space electronic structures. Spectroscopy that couples to charge degrees of freedom alone is therefore insufficient to identify the microscopic nature of the oxygen-hole state, but more fine-grained experimental probes may be difficult to access. Many-body energetic benchmarks provide a critical complementary constraint for resolving competing electronic configurations in oxygen-hole systems.

Finally, the present results establish Na$_{2-x}$Mn$_3$O$_7$ as a stringent benchmark system for competing oxygen-hole localization motifs in correlated oxides. These results demonstrate that intermediate oxygen-hole states stabilized within hybrid DFT may not correspond to stable many-body states when localization, hybridization, and Coulomb interactions compete on comparable energy scales.

\section*{Acknowledgments}

This manuscript has been authored by UT-Battelle, LLC, under contract DE-AC05-00OR22725 with the US Department of Energy (DOE). The US government retains and the publisher, by accepting the article for publication, acknowledges that the US government retains a nonexclusive, paid-up, irrevocable, worldwide license to publish or reproduce the published form of this manuscript, or allow others to do so, for US government purposes. DOE will provide public access to these results of federally sponsored research in accordance with the DOE Public Access Plan (https://www.energy.gov/doe-public-access-plan). K.S. and P.R.C.K were supported by the U.S. Department of Energy, Office of Science, Basic Energy Sciences, Materials Sciences and Engineering Division as part of the Computational Materials Sciences Program and the Center for Predictive Simulation of Functional Materials. Part of the computational work used resources of Expanse at the San Diego Supercomputer Center through allocation MAT230005 from the Advanced Cyberinfrastructure Coordination Ecosystem: Services \& Support (ACCESS) program, which is supported by National Science Foundation Grants No. 2138259, No. 2138286, No. 2138307, No. 2137603, and No. 2138296. Additional computational resources were provided by the Argonne Leadership Computing Facility, a DOE Office of Science User Facility operated.

\bibliographystyle{unsrt}
\bibliography{pap}

\clearpage
\onecolumngrid
\centering{
\large{Supplemental Material}
}
\appendix
\renewcommand{\thesection}{\Roman{section}}
\renewcommand{\thesubsection}{\Alph{subsection}}
\renewcommand{\thefigure}{S\arabic{figure}}
\renewcommand{\theequation}{S\arabic{equation}}
\renewcommand{\thetable}{S\arabic{table}}

%\title{Supplemental Material for: ``Many-Body Destabilization of Intermediate Oxygen-Hole States''}

\author{Anirudh Adavi}
\thanks{These authors contributed equally.}
\affiliation{Department of Materials Science and Engineering,
Massachusetts Institute of Technology, Cambridge, MA 02139, USA}

\author{Kayahan Saritas}
\thanks{These authors contributed equally.}
\affiliation{Materials Sciences and Technology Division,
Oak Ridge National Laboratory, Oak Ridge, TN 37831, USA}

\author{Ming Lei}
\affiliation{Department of Materials Science and Engineering,
Massachusetts Institute of Technology, Cambridge, MA 02139, USA}

\author{Das Pemmaraju}
\affiliation{IBM Research, 555 Bailey Ave, San Jose, CA 95141-1003, USA}

\author{Paul R. C. Kent}
\affiliation{Computational Sciences and Engineering Division,
Oak Ridge National Laboratory, Oak Ridge, TN 37831, USA}

\author{Iwnetim I. Abate}
\affiliation{Department of Materials Science and Engineering,
Massachusetts Institute of Technology, Cambridge, MA 02139, USA}
\email{iabate@mit.edu}

%\maketitle

\section{Computational Details}

\subsection{Fixed-Node Diffusion Monte Carlo}

Fixed-node diffusion Monte Carlo (DMC) calculations were performed using the QMCPACK~\cite{Kim_2018,kent_qmcpack_2020} package together with the Nexus workflow environment. Norm-conserving correlation-consistent effective core potentials (ccECPs)~\cite{bennett_new_2017,bennett_new_2018,kincaid_correlation_2022} were used for Na, Mn, and O atoms. Two-body and three-body Jastrow factors were optimized through variance and energy minimization procedures.

Twist averaging was employed to reduce one-body finite-size effects, while finite-size extrapolation procedures were used to minimize two-body errors arising from periodic image interactions. An imaginary time step of 0.005~Ha$^{-1}$ was used throughout the production DMC calculations.

The DMC total energies are variational with respect to the nodal surface of the trial wavefunction. Trial wavefunctions were therefore generated using several electronic-structure approaches, including HSE and DFT+$U$+$V$ calculations, to assess the sensitivity of the many-body energetics to the underlying approximate electronic structure.

\subsection{Trial-Wavefunction Generation}

Single-determinant Slater-Jastrow trial wavefunctions were generated from Quantum ESPRESSO calculations. For hybrid DFT calculations, the exact-exchange fraction $\alpha$ was varied between 0 and 0.35. For DFT+$U$+$V$ calculations, both onsite Hubbard interactions and inter-site interactions were explored.

The DFT+$U$+$V$ calculations included:
\begin{itemize}
    \item onsite $U_{3d}$ interactions on Mn atoms,
    \item onsite $U_{2p}$ interactions on O atoms,
    \item inter-site $V_{3d\text{-}2p}$ interactions between neighboring Mn and O orbitals,
    \item and inter-site $V_{2p\text{-}2p}$ interactions between oxygen atoms neighboring the Mn vacancy.
\end{itemize}

The $V_{2p\text{-}2p}$ interaction was selectively applied to oxygen atoms surrounding the Mn-vacancy environment because these sites host the oxygen-hole states.

\subsection{Density Extrapolation}

Spin and charge densities obtained from DMC calculations were evaluated using mixed estimators. Pure DMC densities were estimated using the linear extrapolation formula
\begin{equation}
    \rho_0 = 2\rho_{\text{DMC}} - \rho_{\text{VMC}},
    \label{eq:density-extrapolation}
\end{equation}
where $\rho_{\text{DMC}}$ is the mixed DMC density and $\rho_{\text{VMC}}$ is the variational Monte Carlo density.

\section{Hybrid DFT and DFT+$U$+$V$ Benchmark Calculations}

The relative stability of localized and split-polaron configurations was systematically explored across multiple approximate electronic-structure methods.

Hybrid DFT calculations consistently stabilized the split-polaron configuration across the full range of exact-exchange fractions investigated. Increasing $\alpha$ progressively favored the split-polaron state relative to the localized state.

In contrast, DFT+$U$ calculations generally produced substantially more localized oxygen-hole states. Even nominal split-polaron geometries frequently relaxed toward partially localized configurations.

Including inter-site interactions within DFT+$U$+$V$ partially restored the split-polaron character predicted by HSE. In particular, the O-$2p$/O-$2p$ interaction surrounding the Mn-vacancy environment strongly influenced the degree of oxygen-hole delocalization.

\section{Projected and Local Density of States}

\begin{figure}[ht]
    \centering
    \includegraphics[width=0.6\linewidth]{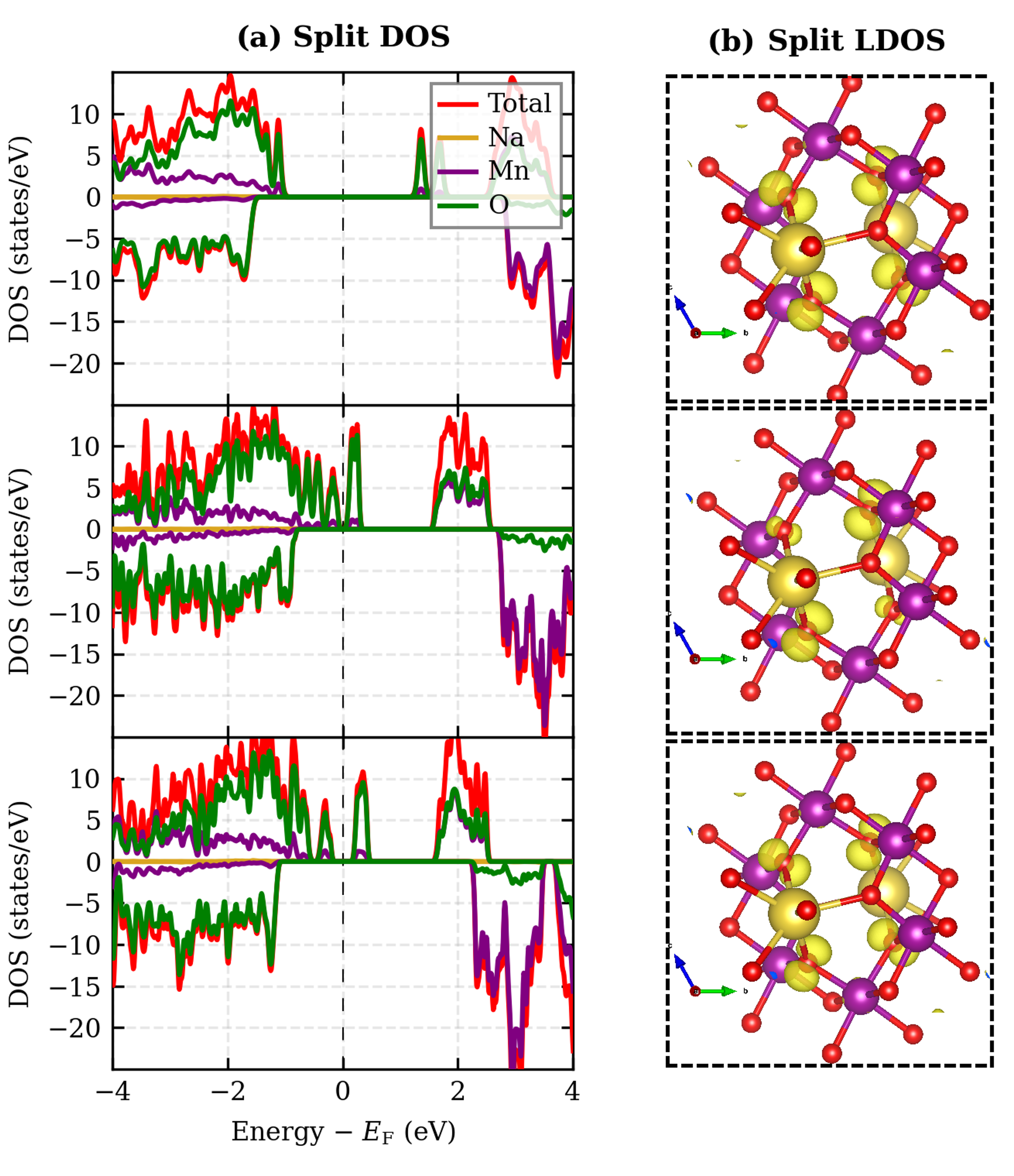}
    \caption{(a) Projected density of states and (b) local density of states for split-polaron configurations obtained from HSE (top), DFT+$U$ (middle), and DFT+$U$+$V$ (bottom) calculations. The oxygen-derived mid-gap states associated with oxygen-hole formation are shown together with the corresponding localization motifs.}
    \label{fig:pdos}
\end{figure}

Fig.~\ref{fig:pdos} shows representative projected density of states (PDOS) and local density of states (LDOS) calculations obtained using HSE, DFT+$U$, and DFT+$U$+$V$ methods.

All methods predict oxygen-derived unoccupied mid-gap states associated with oxygen-hole formation. However, the spatial distribution of the corresponding hole-state density differs substantially among the methods.

Hybrid DFT produces hole-state density distributed across four oxygen atoms neighboring the Mn vacancy, consistent with split-hole polarons. DFT+$U$ calculations instead yield substantially more localized oxygen-hole states. DFT+$U$+$V$ calculations recover partially distributed oxygen-hole states depending on the chosen inter-site interaction parameters.

These results demonstrate that the stability of intermediate oxygen-hole configurations depends sensitively on the balance among localization, hybridization, and Coulomb interactions.

\section{DFT Energy Landscapes Across U and V Parameter Spaces}

\begin{figure}[ht]
    \centering
    \includegraphics[width=\linewidth]{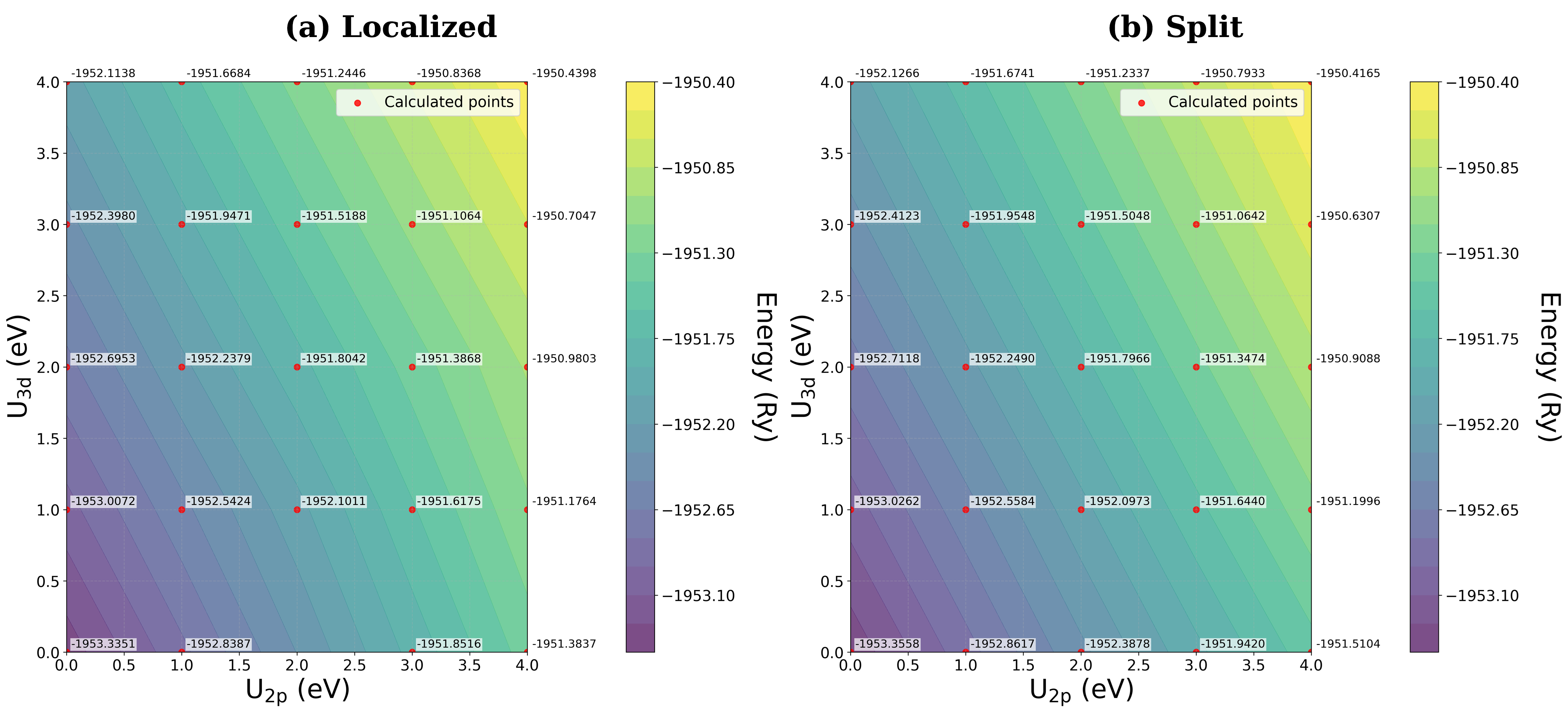}
    \caption{DFT total energies of (a) localized and (b) split polaron across $U$ grid.}
    \label{fig:loc_spt_ene_U}
\end{figure}

Fig.~\ref{fig:loc_spt_ene_U} displays the DFT+$U$ energy landscapes for localized and split-polaron configurations across the explored Hubbard-$U$ parameter space.

The DFT+$U$ energy landscapes show substantial sensitivity to the onsite Hubbard parameters. Increasing $U_{\mathrm{O}\text{-}2p}$ and $U_{\mathrm{Mn}\text{-}3d}$ modifies the relative stability of localized and split configurations and strongly influences the oxygen-hole localization behavior.

\begin{figure}[ht]
    \centering
    \includegraphics[width=\linewidth]{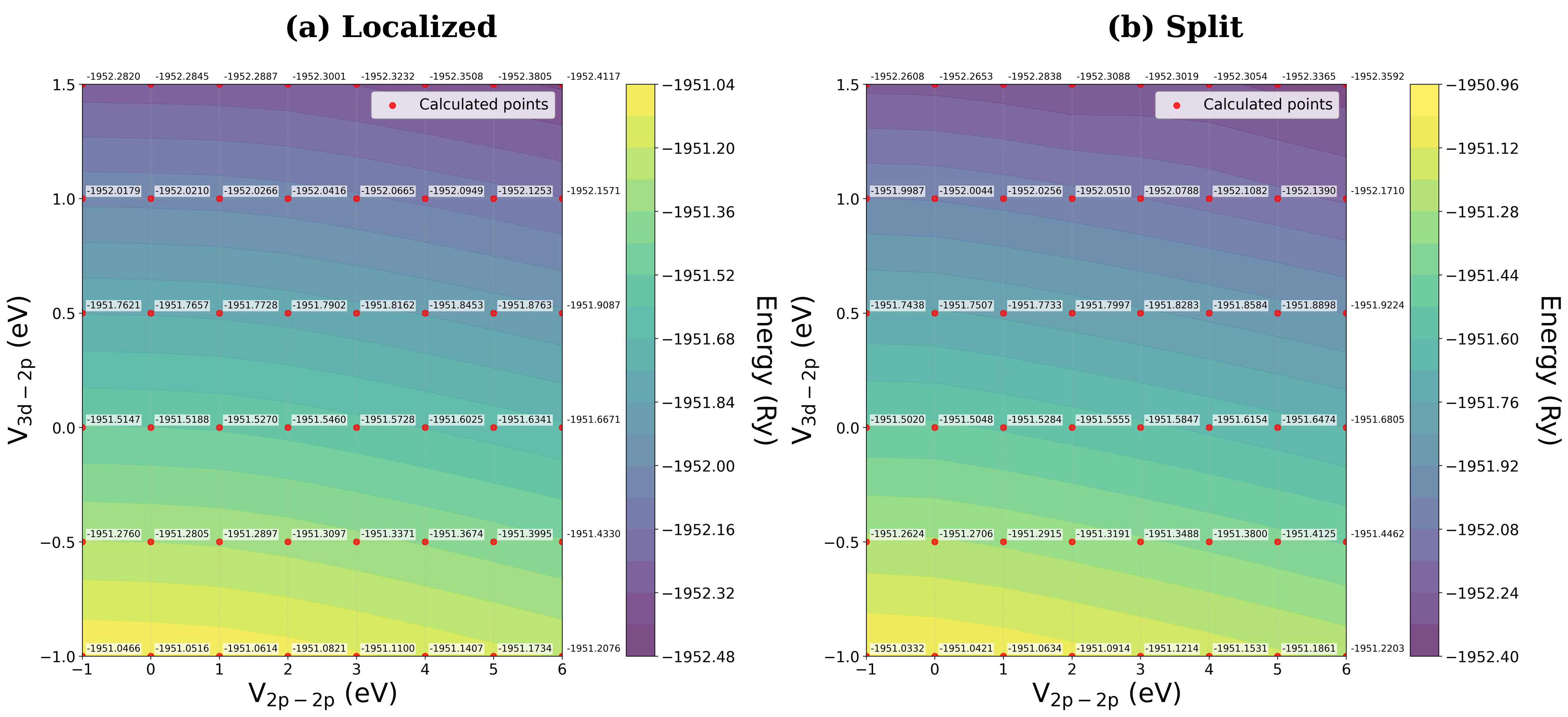}
    \caption{DFT total energies of (a) localized and (b) split polaron across $V$ grid.}
    \label{fig:loc_spt_ene_V}
\end{figure}

Fig.~\ref{fig:loc_spt_ene_V} illustrates the DFT+$U$+$V$ energy landscapes for localized and split-polaron configurations across the explored inter-site interaction parameter space.

The DFT+$U$+$V$ energy landscapes further demonstrate that inter-site interactions, particularly the O-2$p$/O-2$p$ interaction near the Mn-vacancy environment, play an important role in stabilizing partially distributed oxygen-hole states.

Although specific parameter choices can reduce the energetic difference between localized and split configurations, the resulting states remain highly sensitive to the approximate treatment of electron correlation.

\clearpage
\section{DMC Wavefunction Optimization}

DMC wavefunction optimization was performed using both HSE-generated and DFT+$U$+$V$-generated trial wavefunctions. DMC total energies are variational, decreasing as the nodal surface of the wavefunction approaches the exact limit\cite{FOU2001}. The fixed-node approximation is required in DMC calculations, so the nodal surface can only be modified through parameter tuning in the trial wavefunction generation step, which was done through DFT\cite{Kim_2018,FOU2001}. Wavefunction optimization was therefore performed by varying the exact exchange fraction $\alpha$ for hybrid DFT and the U and V parameters for generalized-gradient DFT, using the DMC energy as a benchmark.

\begin{figure}[ht]
    \centering
    \includegraphics[width=\linewidth]{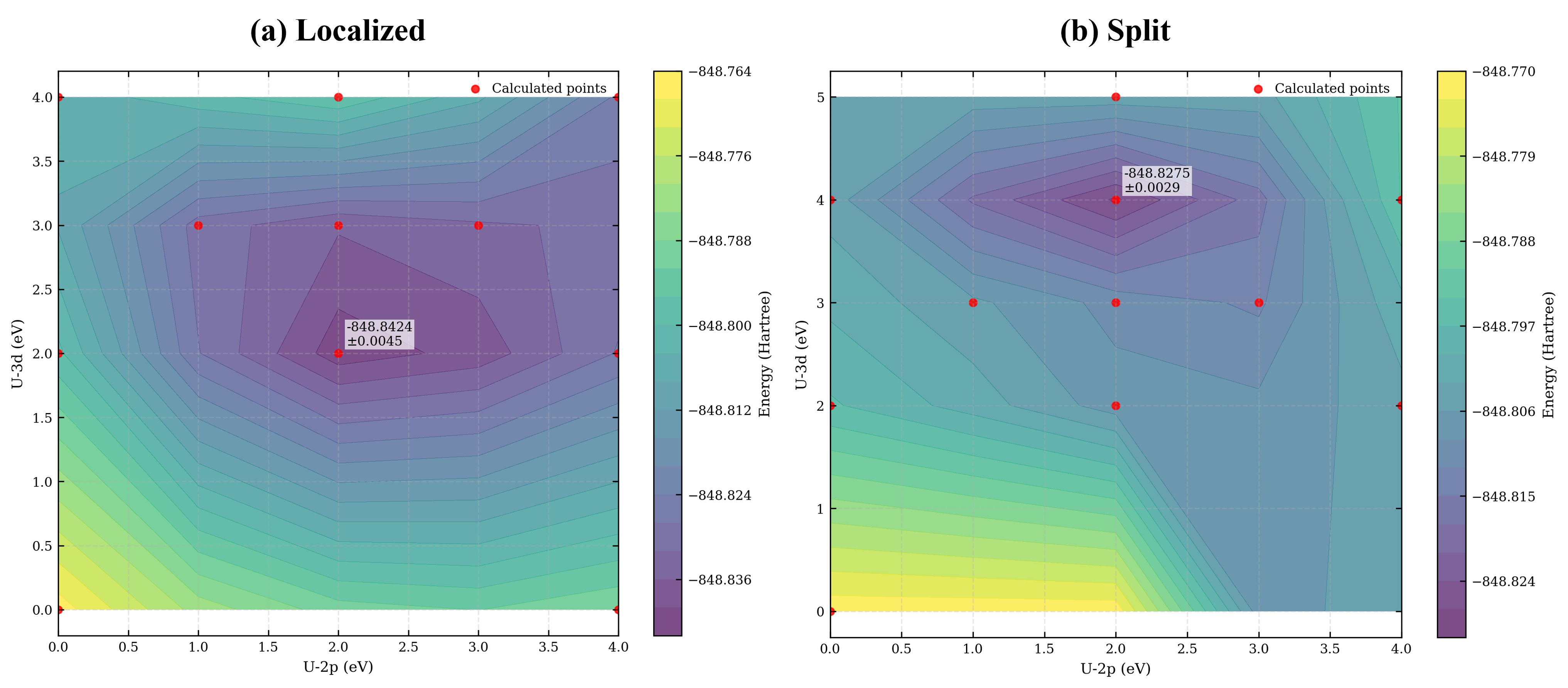}
    \caption{DMC total energies of (a) localized and (b) split polaron geometries across $U$ grid.}
    \label{fig:dmc_u_ene}
\end{figure}

For PBE+$U$+$V$ wavefunction optimization, four parameters ($U_{\mathrm{Mn}\text{-}3d}$, $U_{\mathrm{O}\text{-}2p}$, $V_{\mathrm{Mn}\text{-}3d/\mathrm{O}\text{-}2p}$, $V_{\mathrm{O}\text{-}2p/\mathrm{O}\text{-}2p}$) were optimized. To limit computational cost, the $U$ and $V$ parameters were optimized separately in two independent two-dimensional scans rather than a single four-dimensional scan. After an initial 2D parameter scan across a grid of ($U_{3d}$, $U_{2p}$) values, the $U$ values were fixed at $(U_{3d}, U_{2p}) = (3, 2)$~eV, between the potential wells of the two polaron configurations as seen in Fig. \ref{fig:dmc_u_ene}. These values were then used for the subsequent ($V_{3d\text{-}2p}$, $V_{2p\text{-}2p}$) scans presented in the main text.

For HSE trial states, the DMC energies for both localized and split configurations exhibit minima near $\alpha \approx 0.15$. At this same $\alpha$ value, hybrid DFT predicts the split-polaron state to be lower in energy. DMC instead reverses the energetic ordering and stabilizes the localized oxygen polaron.

DMC calculations based on DFT+$U$+$V$ trial wavefunctions similarly favor the localized configuration, although the energetic separation between localized and split states is somewhat reduced relative to the HSE-generated trial states.

The qualitative many-body preference for localization therefore remains robust across the tested trial-wavefunction classes.

\section{Finite-Size Extrapolation and Time Step Choice}

\begin{figure}[ht]
    \centering
    \includegraphics[width=0.6\linewidth]{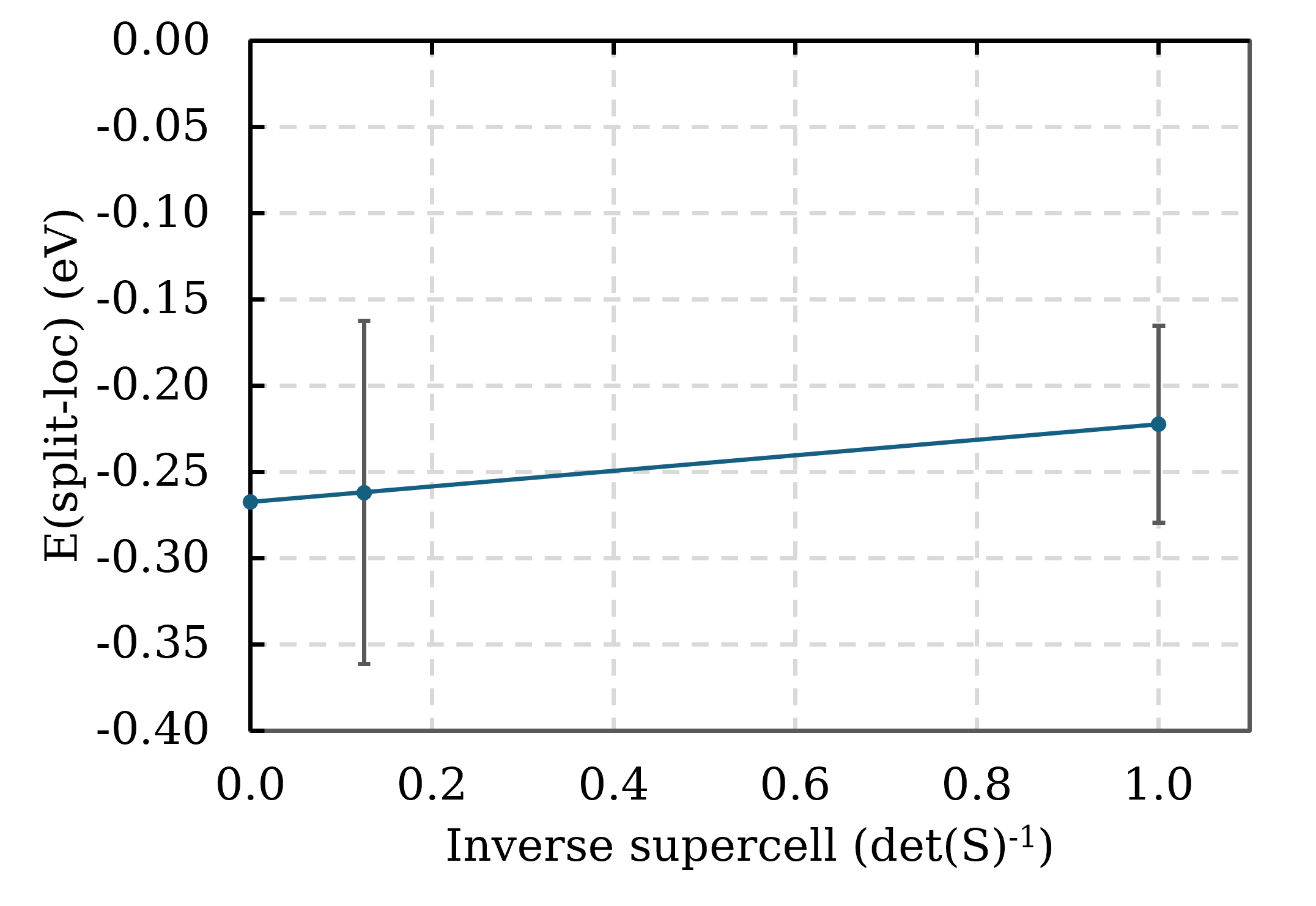}
    \caption{Finite-size extrapolation for difference in configuration energies (split-localized).}
    \label{fig:convergence}
\end{figure}

Fig.~\ref{fig:convergence} illustrates the finite size convergence test. Due to the large size of the unit cell (22 atoms), only a 2$\times$2$\times$2 supercell was tested.

Although the absolute DMC total energies vary slightly with supercell size and twist sampling, the energetic preference remains unchanged throughout the tested parameter space.

The time step was chosen according to prior benchmarking work in the literature on the ccECP pseudopotentials used in this study. A timestep of 0.005 Ha$^{-1}$ is converged within 1 mHa total energy\cite{ANN2020}.

This demonstrates that the many-body destabilization of the split-polaron state is not an artifact of finite-size or timestep effects.

\section{Radial Charge-Density Distributions}

\begin{figure}[ht]
    \centering
    \includegraphics[width=0.9\linewidth]{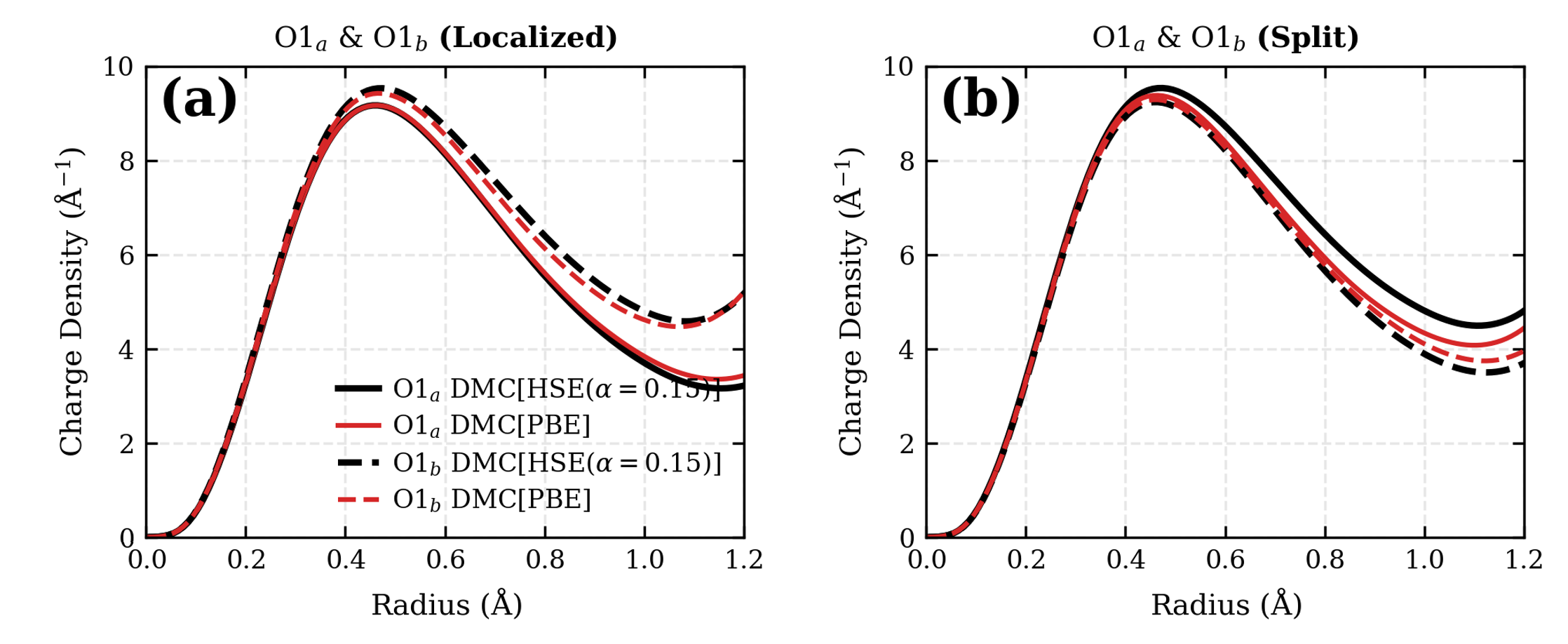}
    \caption{Radial charge density distributions for O1$_\text{a}$, and O1$_\text{b}$ species, shown for both local (loc) and split (split) orbital representations. DMC calculations with HSE trial wavefunctions (DMC[HSE]) are shown in black, while DMC with PBE trial wavefunction are shown in red.}
    \label{fig:radial}
\end{figure}

Fig.~\ref{fig:radial} compares radial charge-density distributions obtained from DFT, VMC, and DMC calculations.

The total charge densities exhibit only relatively modest differences among the various approximate DFT descriptions and the many-body calculations, despite the qualitatively different oxygen-hole localization motifs.

This weak sensitivity of the total charge density helps explain why conventional X-ray absorption spectroscopy cannot uniquely distinguish localized and split-polaron states.

\section{Extended Radial Density Comparisons}

\begin{figure}[ht]
    \centering
    \includegraphics[width=\linewidth]{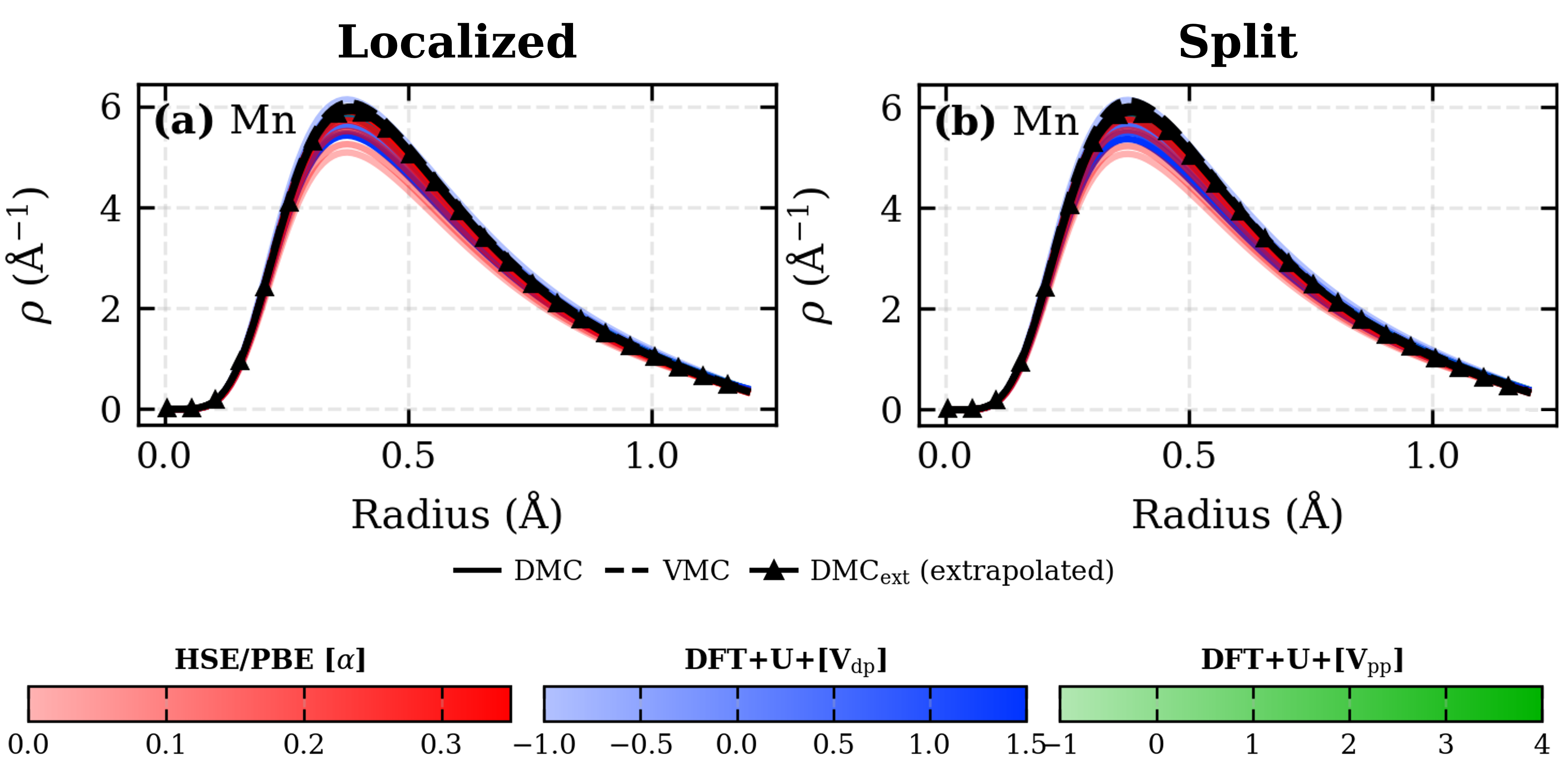}
    \caption{Radial spin density distributions for Mn, shown for both local (loc) and split (split) orbital representations. Each curve represents a different computational method: DMC, VMC, HSE (with varying $\alpha$), and DFT+$U+V$ (with varying $V_{dp}$ and $V_{pp}$ values). These profiles illustrate that the choice of functional and interaction parameters has little effect on the localization and distribution of spin density for Mn species.}
    \label{fig:radial_mn}
\end{figure}

\begin{figure}[ht]
    \centering
    \includegraphics[width=\linewidth]{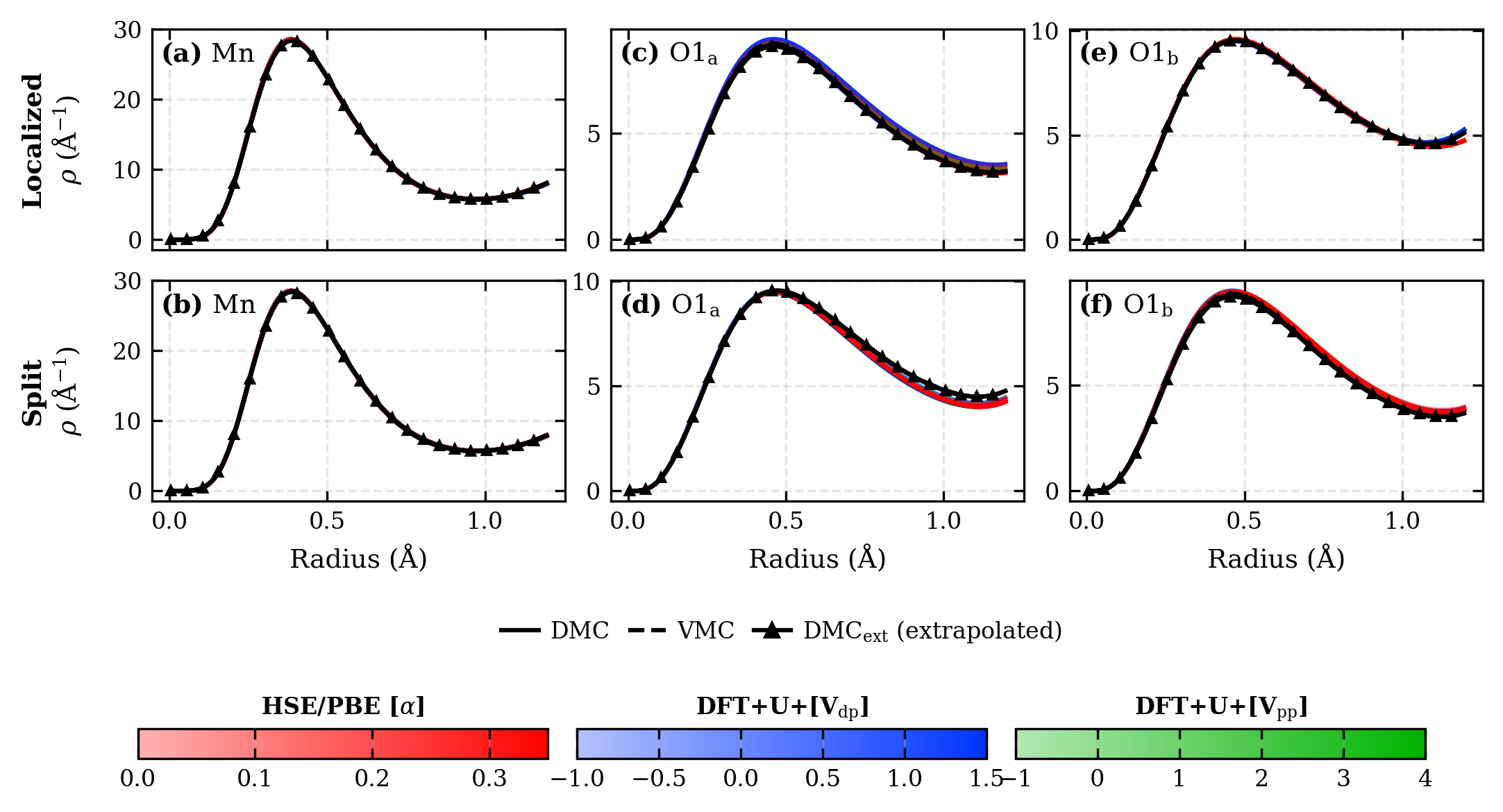}
    \caption{Radial total charge density distributions for Mn, O1a, and O1b species, shown for both local (loc) and split (split) orbital representations. Each curve corresponds to a different electronic structure method: DMC, VMC, HSE (with varying $\alpha$), and DFT+$U+V$ (with varying $V_{dp}$ and $V_{pp}$ values). The $V_{dp}$, $V_{pp}$, and $\alpha$ values are annotated at the peak of each respective curve. The comparison highlights the influence of different functionals and interaction parameters on the spatial distribution of charge around each atomic species.}
    \label{fig:radial_1}
\end{figure}

Fig.~\ref{fig:radial_mn} and Fig.~\ref{fig:radial_1} present extended radial spin-density and total charge density comparisons for Mn, O1$_\text{a}$, and O1$_\text{b}$ sites across HSE, DFT+$U$+$V$, VMC, and DMC calculations.

Spin-density distributions are substantially more sensitive to the oxygen-hole localization character than the total charge density because the oxygen hole corresponds to an unpaired spin degree of freedom. We find that increasing $\alpha$ for HSE calculations or decreasing $V_{dp}$ for the PBE+$U$+$V$ calculations results in greater concentration of spin density around Mn atoms, while the $V_{pp}$ parameter has little effect. However, the variation of parameters in DFT has overall little qualitative impact on the spin density distribution around Mn atoms.

We also find that the total charge densities on Mn, O1a, and O1b species show minimal sensitivity to the variation of $\alpha$, $U$, and $V$ parameters, as illustrated in Fig.~\ref{fig:radial_1}. VMC and DMC calculations also do not significantly alter the distribution of total charge density.

For localized-polaron configurations, the spin density remains strongly concentrated on a single oxygen atom neighboring the Mn vacancy.

For split-polaron configurations, HSE predicts nearly symmetric spin polarization across neighboring oxygen atoms. In contrast, DMC calculations generated from HSE trial wavefunctions strongly enhance the asymmetry between neighboring oxygen sites, producing spin-density distributions that more closely resemble localized oxygen polarons.

DMC calculations based on other trial-wavefunction classes exhibit weaker asymmetry but retain the same overall energetic preference for localization.

\section{Additional Discussion of Trial-Wavefunction Dependence}

The strong sensitivity of the split-polaron state to the trial-wavefunction class suggests enhanced static correlation and near-degeneracy among competing oxygen-hole localization motifs.

Hybrid DFT predicts nearly degenerate oxygen-hole distributions across neighboring oxygen sites, indicating that multiple symmetry-related configurations may lie close in energy. Under such conditions, multideterminant trial wavefunctions or explicit orbital optimization could potentially provide a more balanced description of the correlated many-body state. Nevertheless, within the present single-determinant DMC[HSE] framework, the split-polaron state consistently destabilizes relative to localization. 

\end{document}